# Scientific discourse on YouTube: Motivations for citing research in comments

**Striewski, Sören** — University of Kiel, Germany | stu216065@mail.uni-kiel.de

**Zagovora, Olga** — GESIS – Leibniz Institute for Social Sciences, Germany | olga.zagovora@gesis.org

**Peters, Isabella** — ZBW Leibniz Information Center for Economics, Kiel, Germany | i.peters@zbw.eu

## ABSTRACT
YouTube is a valuable source of user-generated content on a wide range of topics, and it encourages user participation through the use of a comment system. Video content is increasingly addressing scientific topics, and there is evidence that both academics and consumers use video descriptions and video comments to refer to academic research and scientific publications. Because commenting is a discursive behavior, this study will provide insights on why individuals post links to research publications in comments. For this, a qualitative content analysis and iterative coding approach were applied. Furthermore, the reasons for mentioning academic publications in comments were contrasted with the reasons for citing in scholarly works and with reasons for commenting on YouTube. We discovered that the primary motives for sharing research links were (1) providing more insights into the topic and (2) challenging information offered by other commentators.

## KEYWORDS
YouTube comments; Scholarly communication; Altmetrics; Social media metrics; Social media studies

## INTRODUCTION

YouTube is an important source of user-generated content, as well as the world's second most viewed website, behind Google (https://www.statista.com/statistics/1201880/most-visited-websites-worldwide/). In 2021 YouTube's user base contained a total of 2.24 billion registered users (https://www.statista.com/forecasts/1144088/youtube-users-in-the-world) that on average spent "23.2 hours per month streaming videos and interacting with video content via YouTube mobile app" (https://www.statista.com/statistics/1251137/top-video-and-streaming-apps-by-hours-watched-monthly/). Users may engage with videos and comments on the site by posting comments, subscribing, sharing, liking, and disliking content. The amount of engagement with YouTube content is often reflected in quantitative measures, and it has been shown that comments to videos, are strong indicators for reflecting the popularity of a video, instead of the times a video was liked (Chatzopoulou et al., 2010). This has been attributed to the greater effort required to comment on a video than to click the like button (Haustein, 2016). Moreover, evidence has been found that comments serve discursive functions on YouTube and that users utilize them to communicate (Lindgren, 2012). Despite the fact that comments appear to be a rich source for learning more about users' social media behaviors, little research on comments has been conducted thus far. This is most likely due to the enormous volume of available content and the absence of structure, which requires a significant amount of manual labor and advanced processing to be organized. Furthermore, social media platforms are heavily influenced by how users interact with them, which may result in ever-changing types of comments, necessitating ongoing modification of established patterns of analysis (Madden et al., 2013).

The motivations of users to upload and engage with videos on YouTube are manifold, e.g., entertainment, seeking information, seeking self-status (Madden et al., 2013). Recently, it has been demonstrated that a significant percentage of YouTube-content originates from scholars seeking to communicate their findings, even though the great majority of YouTube-videos are unrelated to academia and academic topics (Zagovora & Weller, 2022). However, users are increasingly searching for research results on YouTube to learn more about scientific information, particularly for subjects related to science, technology, and medicine (Allgaier, 2016). Interestingly, videos by researchers or that present scientific topics often address norms that are routine and necessary in the academic world, but not so much in public communication areas, such as newspapers or social media. For example, building the argumentation on and formally citing scientific publications is a required practice in scholarly communication (MacRoberts & MacRoberts, 1987), but newspapers or other public outlets are not required to formally cite studies they report on. However, video producers on YouTube and other social media platforms cite research publications in their videos and frequently include scientific papers in video descriptions (Mehrazar et al., 2019). Furthermore, video commenters refer to scientific articles in their comments, leading us to hypothesis that commenting is a means of having a real discourse or scientific dispute with the video creators and/or other commenters of the video.

In the last decade, in the context of the evaluation of research work, the analysis of occurrences of research papers on social media has been discussed as 'social media metrics' or 'altmetrics' (Priem et al., 2011). Oftentimes, the focus of interest in this field is only laid on the quantifiable information of social media-appearances of research (e.g., the number of times a publication was tweeted on Twitter), which has led to a strong underestimation of the impact research publications have on such platforms. The majority of such altmetrics studies neglected the more qualitative information on how users engage with research results on social media and what can be learned from all types of user-generated content for research evaluation (Haustein & Peters, 2012). The three observations from the preceding paragraphs prompted our research into comments with links to scientific articles on YouTube videos. We aim at exploring why YouTube-users reference scholarly publications in their comments and – in a broader sense – if and why they apply scholarly writing practices to a social media environment and whether those reasons correspond or contradict motivations to cite that are found in the scholarly environment, i.e., in scholarly communication, in which research discuss and exchange research topics among each other.

The contribution of this paper is threefold: We provide a) a categorization schema for comments that include references to research works and b) which is applied to a set of 300 YouTube-comments to exemplify the approach and c) which is compared to other categorization schemas that have been developed to describe reasons to cite articles in scholarly communication. For that, we have adapted the categorization schema from Madden et al. (2013) which provided 10 broad and 58 narrow classes to describe all types of comments, even those that do not contain URLs. In an iterative process we have refined and developed this schema further to fit categories to the particularities of comments, including links to scholarly publications. With our work we can shed light on the different environments in which engagement with research publications take place (academic publishing and scholarly communication as well as social media), which effect they have and how this understanding can provide additional information on forms of user engagement and impact of research. Such knowledge, among other things, can help to inform the area of altmetrics and the creation of meaningful evaluation techniques for research works (Haustein, 2016; Lemke et al., 2020).

**RELATED LITERATURE**

**Commenting on YouTube**
YouTube's content is diverse and international, providing the chance to distribute content to a very broad population of site users who may be on the platform for a variety of reasons. In order to eventually derive a better categorization schema that portrays the genuine purpose of the users' comments that contain URLs to scientific articles, it is necessary to first study the reason for engagement with YouTube videos in the form of comments. There has been very little qualitative research regarding why YouTube viewers engage with videos or other commenters, but the ones that are still relevant to the rapidly developing and constantly changing video platform are summarized in this chapter. Thelwall and Sud (2011) investigated 37,533 comments and discovered that just 0.5 percent of viewers leave a comment, that 72.2 percent of commenters were male, and that the median age of commenters was 25. They also discovered that videos with a high number of comments tend to have disproportionately strong negative sentiments communicated in them; while videos with fewer comments have primarily positive responses. Madden et al. (2013) analyzed 66,637 YouTube user comments to develop a categorization schema on reasons to post comments. They proposed ten broad categories and 58 subcategories, with the main categories being Information, Advice, Impression, Opinion, Responses, Expression of personal feelings, General Conversation, Site processes, Video content description, and non-response comments. This work provided us with a starting point for our own categorization schema.

Khan (2017)'s work is especially important in this sense, since it explained the reasoning for users engaging with each other or with the video. According to the findings of this study, commenting activity is the best predictor of social engagement on YouTube. A typical user may begin or engage in a debate centered on the video's content by submitting comments. They discovered five elements that motivate individuals to interact with user content: Seeking Information; Giving Information; Seeking Self-Status; Social Interaction and Relaxing Entertainment. Furthermore, commenting was indicated by YouTube visit frequency and gender, meaning that males were more inclined than females to comment on YouTube videos, which also fits with the observation from Thelwall and Sud (2012).

Schultes et al. (2013) suggested that all comments could be grouped into three types. Type 1 (T1) includes offensive discussion posts, insubstantial discussion posts, and typical discussion posts. Type 2 (T2) comprises spam or offensive postings, short emotional shout-outs, and insubstantial posts. Type 3 (T3) comprises references to video content, contributions to video content, and normal statements. Their research discovered that the negative image of YouTube comments stems from the high number of T2 comments, but that users continue to utilize comments owing to the added value (information, entertainment, social exchange, etc.) generated from T1- and T3-comments. Even though some categorization schemas are more descriptive or less so, the majority of the categories do have some type of similarity, which will provide us with a starting point for developing our own categorization schema that will help us answer the research question. Furthermore, this is a point of comparison to examine how comments that mention scientific publications differ from comments that do not cite research works.

**YouTube and other video platforms as a source for scientific information, discourse, and altmetrics**

YouTube has been widely used in academic contexts. Burke et al. (2009) investigated YouTube as a source for teaching materials. Kousha et al. (2012) examined the content of videos referenced in scientific publications. The authors discovered a consistent increase in the number of videos linked from scholarly articles. Thelwall et al. (2012) researched videos published by identifiable academics on Twitter. Most of those videos were either used to demonstrate particular scientific phenomena or to present their studies to an academic audience at a conference. Meanwhile, the main purpose of the most popular videos, according to the view counts, was to disseminate information to the public in the form of public lectures or TV shows. Welbourne & Grant (2016) confirmed similar findings, namely, the content created by professional science communicators attracted fewer views and subscriptions than user-generated content. The turnout for niche academic videos is so low that video dissemination cannot be justified by view numbers alone.

According to studies (Allgaier, 2018; Allgaier, 2019; Welbourne & Grant, 2016; Shapiro & Park, 2015), YouTube is utilized as a communication tool in the sciences (e.g., Medicine or Environmental sciences). Shapiro & Park (2015) examined the public's engagement with science-related information and found that regardless of the opinion narrative, 'science-based comments dominated' in the comment's threads under the videos. These findings provide us with a new source of information, namely comments, where scholarly communication involving the public occurs. YouTube has been included as a source of altmetrics in most altmetrics aggregators. Metrics obtained from the YouTube platform can reveal the broader impact of research on society (Bar-Ilan, 2018; Shapiro & Park, 2015; Zagovora & Weller, 2018). For example, Sugimoto et al. (2013) found that the proportion of YouTube likes is positively correlated with the Web of Science publication counts of the presenters (i.e., scientists in this case). These findings have been revealed for videos from TED Talks series, that disseminate science and technology. Although presenting a TED Talk does not appear to affect later citation counts for researchers, it has been demonstrated that academic speakers outperformed the average for their area in terms of citation impact (Sugimoto et al., 2013). These findings are consistent with those of Sugimoto & Thelwall (2013), who discovered that the number of views and comments on TED videos had a weak positive correlation with Google Scholar citation counts. These evidences imply that YouTube videos and their associated popularity numbers might be used as an indicator of research impact or broader social impact.

Despite the potential of comment threads and some popularity indicators, YouTube continues to play a limited part in scholarly communication and altmetrics research, in contrast to numerous studies of Twitter or Mendeley. This might be due to the overall underrepresentation of YouTube studies in social media research, as well as a lack of established theories and methods to address the complexity of YouTube as a platform, as pointed out by Allgaier (2018). With this study, we will try to obtain more details about YouTube comments and whether referencing research publications have been used for similar reasons as in scholarly communication.

**Citation motivation in scholarly communication**

Citations serve various functions in academic communication: they link and distribute research publications, they help in the exploration of research fields (e.g., by browsing references), and they are used to evaluate research. (i.e., bibliometrics or scientometrics; Di Iorio, 2013). Also, citations are manifestations of one of the major principles in scholarly work and publishing: "give credit where credit is due" (MacRoberts & MacRoberts, 1987). Although referencing is a rather organized and consistent process across diverse academic fields, citation intentions are heavily influenced by factors such as disciplinary customs. (Hernández & Gomez, 2015). Hence, most of the studies and approaches to identify citation motivations use forms of intellectual coding to address this issue, although this is a time-consuming and laborious task. Hernández and Gomez (2015) have reviewed eleven categorizations of citation functions and found that they greatly vary in their granularity (from 2 to 35 categories). Di Iorio et al. (2013) and Peroni and Shotton (2012) have introduced CiTO, the Citation Typing Ontology, which includes more than 30 types of citation motivations. Because citations may be discovered using cue words (Jörg, 2008), the authors employ full sentences and included verbs to infer the underlying citation motivation using automated approaches. In their study, they have found that 'Cites as information' is the most commonly assigned citation motivation; however, this class was also assigned if no other category was matching.

Teufel et al. (2006) have introduced 12 classes that not only incorporate the reason to cite, but that also incorporate the sentiment/polarity of the citation, e.g., whether it is neutral, negative or positive. They discovered that differentiating between neutral and weak (which points out a weakness in cited work), is challenging because authors frequently conceal their citation motives, for example, to be polite or to meet the norms of their scientific community. Cronin (1981) and Hernández and Gomez (2015) arrive at similar conclusions. It has been proposed that future automatic assignments of citation motives take into consideration the argumentative structure of scientific publications and draw motivations from the citation's placement in the text, e.g., introduction or related work section (Di Iorio et al., 2013). Challenges for manual and automatic approaches to assign motivations to citations include identification of an appropriate citation context window and the detection of implicit citations (Di Iorio et al., 2013). In contrast to formal scholarly publications, comments to YouTube videos display no argumentative or discursive structure – with the exception of the tree-like structure that develops when commentators start a thread under one of the comments. Also, there are no community standards or other affordances on YouTube that may produce forms of structured comments – except for guidelines that ban the use of swear words; not even full sentences are required in a comment. Because of that and to fully explore the various forms of commenting, we have conducted intellectual coding of comments.

## DATASET & METHOD

In the following we will describe how the dataset of comments including references to scientific publications was created and processed and how the categorization schema was developed and applied.

### Dataset creation

The initial data consisted of 4,336,304 comments from 15,613 distinct YouTube videos (as of December 2017) that were referencing research publications in the video description section. Data about YouTube videos have been provided by Altmetric.com (https://www.altmetric.com). All the comments have been mined using the YouTube API. First, all URL links from each comment were extracted. Second, the retrieved URLs were then divided into four categories: research publications and preprints, scientific news, mixed links, and others. Only links from the research publications and preprints category were used for the purpose of this study.

### Mining links to research publication and preprints

A two step-process was carried out to mine links to research publications and preprints: 1) We extracted document identifiers (PMC, DOI, PMID, Handle, ISBN, ISSN and arXiv) from comments by using regular expressions, and we marked them as research publications and preprints. 2) From the remaining comments with URLs, we created a list of common domain names utilized in URLs. Then two researchers manually checked the top 500 most frequent domains and tested if those might host research publications or preprints (e.g., Nature, ScienceDirect, Ncbi.nlm.nih.gov). Via this approach, 1,691 more links were identified as potential links to research publications or preprints. Since some web pages of the same domain names might lend to non-research articles (http://www.sciencemag.org/news/), we deleted them from the dataset using manually created heuristics.

### Dataset Cleaning

The initial dataset of comments with links to research publications and preprints consisted of 5,730 URLs referenced in a comment. The cleaning procedure was carried out to arrive at URLs that contained the bare minimum of information required to access the article or preprint on the web, as well as to delete URLs that did not lead to articles or preprints. So, the protocols of all URLs were standardized in the first stage. Following that, a more sophisticated normalization approach was applied to locate URLs that link to the same article but varied by a set of characters that may be omitted. The removal of characters may affect the semantics for certain URLs, therefore every change was carefully evaluated for semantic preservation.

| Persistent identifier | Number of URLs | Number of distinct URLs | Number of distinct comments |
|---|---|---|---|
| DOI | 538 | 429 | 455 |
| ISSN | 0 | 0 | 0 |
| ISBN | 3 | 3 | 3 |
| Arxiv | 188 | 134 | 153 |
| PMC | 848 | 620 | 672 |
| PMID | 2,380 | 1,564 | 1,006 |
| Handle | 12 | 12 | 12 |
| Others | 1,415 | 1,068 | 1,058 |

Table 1: Number of links and distinct links for each persistent identifier of research publications and preprints.

Following the normalization process, the availability of each URL was tested by opening each URL with a http get request and then filtered. Next, URLs that were manually identified as not leading to scientific articles or preprints were removed from the dataset (URLs that just lead to homepages or search engines). In addition, every URL was manually reviewed numerous times to arrive at the final cleaned dataset with 5,368 URLs. Table 1 represents the number of total and distinct URLs for each persistent identifier after the cleaning process.

**Development and description of categorization schema for comments**

An iterative strategy of building and applying the categorization schema was used to identify the reasons why individuals mention scholarly publications in comments to YouTube videos. This was accomplished by analyzing the dataset, finding potential categories, and then testing these on further data to determine if they could be used in their present form; hence, we applied a qualitative analysis approach. Madden et al. (2013)'s categorization approach was used as a starting point to categorize a random sample of 300 comments from the cleaned dataset. This sample was extracted with Java and was based on the number of distinct comments rather than the number of lines, because multiple publications might be cited in a single comment, resulting in multiple lines in the document for a single distinct comment. Furthermore, only comments in English and videos that were still publicly available at the time of study (February 2022) were considered in order to appropriately categorize them.

Madden et al. (2013) categorized user comments to develop a categorization schema that can be used to identify all types of YouTube comments, even those that do not reference any scientific publications. Also, they allowed assignment of comments to multiple categories, indicating that the categories are not selective enough for our research focus. Hence, the 10 broad categories and 58 subcategories were manually examined and reduced to only those judged relevant to the research question. The first step in the process of finding the most fitting categories included reading all the 300 randomly selected comments from the cleaned data set to get a feeling of how those comments are structured and how the commenters are interacting with each other. The comments were looked up in the browser, by finding the video using the 'videoID' in the cleaned dataset.

With the newly discovered insight into the composition of comments and their relationship to other commenters, the possible categories from Madden et al. (2013) were then iteratively tested against the comments, by recording all the categories into which a single comment could be categorized. After categorizing all the comments in this manner, it became evident that several categories may be merged or removed, either because they did not occur frequently enough or were not included in the dataset. For example, because no comment indicated anything regarding the site's operations, such as video posting, this category was omitted. At the same time, comments that did not fit any of the categories from Madden et al. (2013) were assigned newly created categories, for example 'self-advertisement'.

This procedure resulted in the creation of our initial categorization schema, partly adapted from Madden et al. (2013) and partly extended with our own categories. The primary categories are 'Information', 'Response', and 'Other', with varying amounts of subcategories. For the 'Information' category, the subcategories 'Request', 'Give', and 'Suggest' were chosen. The 'Response' category can be refined with the subcategories 'Agree', 'Disagree', and 'Challenge'. The 'Other' Category was created to identify comments that lack context or any other information besides the URL to the scientific publication. All 300 comments were recategorized using this first-generation schema. After categorizing the first 30 comments (10% of the data), the categories 'Response' and 'Information' were combined since they were difficult to distinguish and many comments fell into both categories. Instead, 'Challenge' was designated as a primary category to represent comments that not only impart information, but also directly oppose another person's claim. The updated schema was used to categorize a total of 150 comments (50 percent of the data), and then another reevaluation phase was performed. The 'Challenge' category was further subdivided based on who is being referred to, whether it is the video author or another commenter, and if the statement in the comment is intended to be insulting. Furthermore, the 'Other' category was expanded with the subcategories 'Video Citation' and 'Self-advertisement.' It was also investigated whether certain categories have different types of sentiment, so that additional categories for the sentiment of providing information might be needed to be established. However, nothing noteworthy was discovered here because most of the comments with a negative sentiment are included in the 'Challenge' category and there were too few comments with a highly positive sentiment, meaning that the commenter is praising someone else, that the introduction of a new subcategory would have been reasonable. Following this round of categorization, a review session took place with the other authors of this paper (OZ, IP) who were not engaged in the construction of the schema. Both categorized 100 identical comments, and between these three a Fleiss' kappa-agreement score of 0.54 was reached. This moderate agreement score could be improved with minor tweaks, since some categories were still too ambiguous.

Following this review session and collaborative discourse of the authors, the 'Information' subcategories were divided into three primary categories (Provide, Suggest, Request) to better reflect the commenters' intention. The 'Insult' subcategory was deleted since there were not enough comments categorized as such (which might be due to YouTube deleting comments containing any swear words). The remaining subcategories, 'Challenge person' and 'Challenge video content', were moved to the newly established 'Provide' category (see Table 2 for the final categorization scheme). Even though we attempted to make the categories as distinct as possible, they are not mutually exclusive because comments may contain content that fits into more than one category. As a result, the following hierarchical system was developed to categorize comments more firmly, as certain categories should overshadow others: Challenge > Provide Information > Request > Suggest > Other.

This revised schema was then used to categorize all 300 comments again, followed by another review session with one of the authors (OZ) and another researcher who is not affiliated with this research and had not previously been part of the categorization process. The other author of this work (OZ) and the researcher both categorized the same 50 comments. As a consequence, a considerable agreement Fleiss' kappa-score of 0.75 was determined between those three, which was a better result than previously obtained and proved that the categorization schema performed effectively in practice. Table 3 shows the outcome of the exemplary coding of 300 comments and the distribution of categories. The full data can be accessed in Striewski et al. (2022).

| Category | Description | Example |
| --- | --- | --- |
| **Provide** | The provide category was adapted from the 'Information' category from Madden et al. (2013) and defined as "Information comments are those that provide factual information about something in the video content, video context or a completely unrelated topic." | |
| - Challenge person | This subcategory is similar to the 'Additional Insight' subcategory, but this category can have a negative sentiment that sometimes can border into insults. The most important difference is that this is always in direct response to a person, and always refutes a claim that was made by another commenter or tries to prove them wrong. | "+Derek Billington No it reduces insulin sensitivity. http://www.ncbi.nlm.nih.gov/pubmed/25475435" |
| - Challenge video content | This subcategory is similar to the 'Additional Insight' subcategory, but this category can be a negative sentiment that sometimes can border into insults. The most important difference here is that this is always in direct response to the video, and always refutes a claim that was made by the specific Youtuber in the video. | "That egg smoking study has been debunked. Several researchers have shared their comments: http://www.ncbi.nlm.nih.gov/pubmed/23177013. As well if I recall correctly the researchers reevaluated their data and found health benefits in egg consumers. Please annotate your video to reflect this. " It would be doing your fans a disservice" if you didn't." |
| - Additional insight | These comments will include not just a link to a scientific paper with a call-to-action, like in the case of the suggestion, but also additional facts or explanations. The important part here is that they provide additional explanations with a neutral or positive sentiment and are not refuting a point. In some cases, this is a response to an Information Request. | "High temperature improves immune cell response and changes their behavior. https://www.ncbi.nlm.nih.gov/pmc/articles/PMC4786079/" |
| **Request** | These types of comments will seek for other people's opinions on the subject or request explanation on a subject, by first stating their current state of knowledge. | "What do you think of this? http://www.ncbi.nlm.nih.gov/pubmed/24865576" |
| **Suggest** | Commentators in this category will offer further reading material on a subject with a call-to-action (to click on the link, which might be indirect or weakly mentioned), but will not provide further explanation. | "This study may be of interest to you: https://pdfs.semanticscholar.org/3924/e479a5152a196d71a6d24dd7082a54bf8a9f.pdf" |

| **Other** | This category comprises any comments that could not be assigned to one of the previous categories. | |
|---|---|---|
| - No context | A comment that only contains a link or the title of the paper linked and the link to the paper. As well as, comments with links where the relation/motiv/intention of the comment cannot be discerned without making assumptions. | "Free markets! https://mises.org/library/science-technology-and-government-0" |
| - Video Description and References | The user who submitted the video provides references in the form of a comment, which might be a direct copy of the description or a portion of it. | "> Population dynamics in France made visible through analysis of mobile phone usage. Credit: Catherine Linard From: http://phys.org/news/2014-10-cellphone-population-density.html Full paper: http://www.pnas.org/content/early/2014/10/23/1408439111.full.pdf+html via +Tiago Peixoto Population dynamics in France made visible through analysis of mobile phone usage." |
| - Self-Advertisement | These comments are intended to inform others about the publication of a scientific article or preprint by the commenter, or to sell a product by making it appear more legitimate by quoting scientific papers. | "We published a perspective on this topic, see R.A. van Santen et al., Phys. Chem. Chem. Phys. 2013, 15, 17038-17063 (http://dx.doi.org/10.1039/c3cp52506f)." |

Table 2. Categorization schema with descriptions and examples.

## RESULTS

In this section we present the results of our study: a) the example application of the categorization schema on 300 random comments and its implications, and b) the comparison of our findings with general motivations to comment on YouTube and with the reasons to cite articles in scientific works.

**Categorization of comments that include links to scientific works**

There are several reasons why people include scientific papers in their YouTube comments, just as there are numerous reasons why people interact with a video (Khan, 2017). Overall, all comments seem to contribute some type of information; the intent behind them, however, varies. The most evident reason to add links to research publications is 'Provide' (Table 3) which represents information comments that provide more factual information about some aspect of the topic of the video. We distinguished between comments that 1) challenge previous commenter(s) in addition to providing information ('Challenge person'), 2) challenge information or some facts provided in the video ('Challenge video content'), and 3) simply provide 'Additional Insights' without arguing or disagreeing with statements provided by previous commenters or statements from the video content.

The subcategory 'Additional Insights' comprises primarily neutral comments, however on rare cases, the sentiment of the comment is positive. That is, the commenter is praising the preceding commenter for their discoveries. Nonetheless, this did not occur frequently enough to warrant the development of a new subcategory. This category may be easily assessed by determining whether they are only offering further insights into the issue in an objective way or are explicitly challenging another commenter's perspective.

Unlike 'Additional Insights', both 'Challenge' subcategories (that dispute other points of view) had higher negative sentiments. Strong words or indirect insults that belittle the other person are more prevalent in this category than in the others. In our dataset, the 'Challenge' subcategories are the most common motivation to include URLs to scientific articles in the comments. This conclusion is consistent with the findings of Schultes et al. (2013), who suggested that YouTube comments have a negative image. This phenomenon can be explained by the option of anonymity for YouTube-users, which may attract more negative comments. However, the comments that attempted to challenge someone were also frequently objective and attempted to describe the scientific publication and why the other person is wrong. Still there were some outliers who attempted to demonstrate their superior 'Intelligence' through scientific publications, they were far fewer than those who added value to the conversation.

In some circumstances, the individual commenting did not want to contribute information, but rather wanted to be informed (category 'Request'; Table 2), therefore they stated their present level of expertise and then asked other users to answer their questions for them. This frequently caused other users to either challenge this presented level of expertise or supply them with further insights, implying that the 'Provide' or 'Suggest' category will nearly always follow after the 'Request' category, with the 'Provide' category being more common. However, the sample size of the 'Request' comments was simply too small to explore this theory further. In the 'Suggest' category, the comments that just provide links to further reading resources without specifically discussing the substance of those reading materials are grouped. The essential aspect here is that there should be a clear call-to-action, such as 'Click on this link' or 'You may read more here.'

The 'Other' category, which is the second-strongest category overall, encompasses three other subcategories namely: 'No context', 'Video Description' and 'Self-Advertisement.' The 'No Contex' comprises of comments that just include the URL to a scientific article without any explanation or other discursive function. Hence, no detailed analysis of the motivations for such comments is possible. The 'No context' and 'Video Description' categories can be explained by YouTube's affordances and community rules that allow for this kind of interaction and content provision. The third category, dubbed 'Self-Advertisement', is an intriguing subcategory that did not appear very frequently. In this category, people actively promoted their own research or attempted to market a product by bolstering its credibility using scholarly papers. One person, for example, attempted to market their own fitness programs by claiming that it was backed up by research.

| Category | Number of occurrences | Percentage in ratio to main category, % | Percentage overall, % |
| --- | --- | --- | --- |
| **Provide** | 199 | | 66,33 |
|     Additional Insight | 93 | 46,73 | 31,00 |
|     Challenge Person | 68 | 34,17 | 22,67 |
|     Challenge Video Content | 38 | 19,10 | 12,67 |
| **Request** | 13 | | 4,33 |
| **Suggest** | 41 | | 13,67 |
| **Other** | 47 | | 15,67 |
|     No Context | 30 | 63,83 | 10,00 |
|     Video Description | 14 | 29,79 | 4,67 |
|     Self-Advertisement | 3 | 6,38 | 1,00 |

Table 3. Percentage of category occurrences in the dataset of 300 comments referencing research publications.

To summarize, most comments (more than 66%) are categorized into two types: to inform (Table 3, 'Additional Insight') and to make a point (Table 3, 'Challenge person' and 'Challenge video content'). It has been shown that incorporating scientific publications in an argument increases the chance to appear scientific and that this can boost persuasiveness (Tal & Wansink, 2014), even if the cues are as simple as a link being provided. This also makes the comment more like a scientific argument, which employs evidence and data to support a claim rather than belief or opinion, because evidence and data can be scientifically reexamined and retested, but beliefs and opinions cannot.

**Comparison of commenting motivations in comments with or without URLs to academic publications**

Another consideration for this work is a comparison of commenting motivations from our schema with generic YouTube comments that do not cite academic publications. This may provide some more information about what distinguishes those two classes of comments – hence, we compare our categorization scheme to those of Madden et al. (2013) and Khan (2017). Madden et al. (2013) is one of the most influential works for this study, and it has one of the most detailed categorization methods for YouTube comments. As previously stated in the Methods section, several categories were excluded in order to establish our own categorization method. Our final categorization approach includes several categories that are quite similar. For example, the 'Provide' category is comparable to the information category since it was developed from it, but it may also suit the 'Opinion' and 'Responses' categories. All our categories may be claimed to match the Madden et al. (2013) categories that give some type of information. However, categories based on fundamental small talk, such as 'General conversation', 'Site processes', and 'Non-response comments', were not present in this dataset and thus do not match any category in our schema.

In addition to that, Khan (2017) discovered five major motivations for people to interact with videos in the comment area, which are as follows: 'Seeking Information', 'Giving Information', 'Self-Status Seeking', 'Social Interaction', and 'Relaxing Entertainment.' Again, 'Seeking and Giving Information' are the key components of comments with URLs to scientific papers in them, as shown in Table 3. However, Self-Status Seeking was not explored in this study, but it stands to reason that at least a percentage of people who try to refute another person's claim in the 'Challenge' category will do so in order to prove their own superior knowledge. This leaves us with categories centered on small chat, namely 'Social Interaction' and 'Relaxing Entertainment', which were yet again not observed in our dataset. In comparison to those two studies and together with Table 3, it is clear that comments including URLs to scientific papers are concentrated on information flow and are rarely driven by small talk, but rather to have a conversation with diverse sentiment linked to it.

**Comparison with citation motivations in scholarly communication**

As previously stated, citations in scholarly communication serve many functions, the most important of which are: distribution of research papers, exploration of research domains, and evaluation of research. Also, citations are a form of reputation building and credit giving (MacRoberts & MacRoberts, 1987). Using our categorization schema, we discovered that the reasons for citing in YouTube comments are not completely dissimilar to the reasons for citing in scholarly communication. 'Provide', for example, has striking resemblance to what was referred to as 'cite as information' in scholarly citation motivations (Di Iorio et al., 2013).

Also, challenging findings of authors or statements of authorities is rather common practice in scholarly communication (also often in the review process). According to CiTO, 'corrects' motivation that corrects statements, ideas or conclusions presented in the cited entity and 'disagrees with' motivation that disagrees with statements, ideas or conclusions presented in the cited entity, are the most similar to our both 'Challenge' subcategories. In our study, suggesting further reading material linked with a call for action could be matched with 'extends' and 'gives background to' motivations by Di Orio et al. (2013) that extend facts, ideas or understandings presented in the cited entity or provides background information for the citing entity.

In our dataset, the 'Self-Advertisement' subcategory represents promotions of (1) some external products as well as (2) the results of scientific investigations per se. In academic literature, both of those cases can be represented in the form of footnote notations where some source would be acknowledged. Inclusion of these sources into the references list would be rather uncommon. However, King et al. (2017) found that self-citations by some academics might be some form of self-advertisement or promotion. In some cases, academics cite their own publications, even if alternative publications (conveying the same utility) from their peers exist. From the altmetrics perspective, self-promotion is not a rare case. According to Ferreira et al. (2021), academics on Twitter self-tweeted two of their own publications on average, accounting for around 27 percent of their authored publications on the platform.

However, the coding has revealed three peculiarities that do not have an equivalent in scholarly citation practices or which would require at least some "creative" reinterpretation of some of them: 'Request', 'Video description', and 'No context.' Actively seeking information about a certain article ('Request') makes use of YouTube comments' discursive character and has no equivalence in formal scholarly publication - unless one considers calls for future work and studies as a form of 'Request'. In the subcategory 'Video Description' the reference list is delegated to the comment in the 'Video Description' – which might be considered similar to the use of footnotes or endnotes in some disciplines. The subcategory 'No context' is only feasible due to the nature of YouTube comments and comment threads, which do not need (con)text in which the reference is placed. Thus, both 'Video description' and 'No context' are rather artifacts of the platform with non-identifiable referencing reasons, than some unique motivations for citing research.

**DISCUSSION & CONCLUSIONS**

Our research attempted to uncover the processes by which YouTube users, as a sample of the general public, interact with scientific topics and publications, as well as how their behavior differs or aligns with citation and commenting practices found in scholarly communication and in YouTube commenting behavior. For that, we have developed a categorization schema that caters to the particularities of comments including URLs to scientific publications and preprints and which has been applied to a random sample of 300 comments.

The developed schema consists of four primary categories and six subcategories, created through an iterative qualitative analysis approach utilizing Madden et al. (2013)'s categories as a starting point. The primary difference between Madden et al. (2013) and our proposed categorization schema is that theirs is a general-purpose approach that can be used to categorize any comments, not simply those that contain links to scientific publications. As a result, comments may be assigned to many categories, making categorization difficult. We developed a schema for more exact categorization while focusing on comments that solely contain links to scientific publications.

The exemplary application of the categorization schema has shown that there are several reasons why users reference scientific publications in their YouTube comments, just as there are various reasons why people engage with a video (Khan, 2017). The 'Provide' category and its subcategories could be assigned to the majority of comments in our sample, accounting for two-thirds of all comments. The 'Challenge person' (almost 23 % of all comments) and the 'Challenge video content' (nearly 13 % of all comments) subcategories together make up the majority of the 'Provide' category, closely followed by the 'Additional Insights' subcategory (31 percent of all comments). The second-largest primary category is 'Other' (almost 16% of all comments) comprises the subcategories 'No context', 'Video Description', and 'Self-advertisement.' Almost 14% of all comments are considered 'Suggest' which highlights the comments and commenters' roles in providing information to the audience and in helping to embed scientific results into related literature but also into the public discourse, i.e., the discourse among YouTube users.

One limitation of our study concerns the random sample of only 300 comments to which the categorization schema was applied. The other limitation is that the comments were mined not from the full scope of YouTube videos but rather from the set of videos that reference research publications in the description section. It is possible that we have missed important and frequent types of comments, also in comments to videos that do not mention research publications in their description, so that our categorization schema might be incomplete. Also, the categorization schema was inductively developed and further refined on basis of the same comments; which is common in approaches of qualitative content analysis. Hence, all comments have found a respective category; which could have been different with a theory-driven deductive approach. Furthermore, the development of a categorization schema is not trivial and may display a certain level of subjectivity (a limitation which has been highlighted by Madden et al. (2013) as well). However, given that we have conducted two runs of categorizations with coders that were not involved in the development of the schema, we are confident that the application of the categorization shows a high amount of validity. A similar limitation derives from the content taken into account for the interpretation and categorization of the comments. Comments were categorized by only considering the video description and the preceding comment, to which the comment under examination is a response. An examination of the tree-structure of comments may potentially yield useful insights, which will be saved for future work. Watching the video might lead to more accurate interpretations of comments (Madden et al., 2013). Additionally, our approach of assigning only one category to the comment may have led to an underestimation of motivations, since a large share of comments could have been described with multiple categories as well. The other limitation is that categorization was conducted on a relatively small sample, which might influence the generalizability. A replication of the study with a larger sample is considered.

The comparison of reasons for referencing scholarly papers in comments to YouTube videos and in scholarly communication has shown that motivations are strongly affected by the environment in which they take place. Norms in and affordances of those environments are key in this regard, since they – to a large extent – prescribe how users carry out communicative practices. However, users can find ways to utilize affordances in a manner not intended by the providers of the platforms and environments. Cronin (1981) has also highlighted that, although common practice in scholarly communication, citing is a highly subjective, complex and strategic act from the author and "[f]ull knowledge of the reasons why an author cites would require omniscience on the reader's part" (p. 17). Assigned categories are necessarily partly speculative about the motivations for citing a particular article.

Despite the presence of platform-specific commenting motivations in the developed schema, the vast majority of comments has shown similar citing reasons as in scholarly communication. In scholarly communication, usually researchers communicate between each other, here we observed YouTube users (who are not necessarily researchers) that mimic or even do use research practices in their comments. Therefore, YouTube comments' threads that include references to research publications should be considered as a new source of 'beyond scholars'-scholarly communication. Moreover, YouTube comments should be tracked and defined by every altmetrics aggregator as a new and separate metric (beyond mentions in video descriptions).

Mai (2011) and Madden et al. (2013) emphasize the role of the medium itself and the (social) contexts in which it is used when applying content analysis and categorization schemas to user-generated content. A certain understanding of the medium and contexts is needed to arrive at plausible interpretations. In our study, we used an interpretation approach to deduce user motivations for referencing articles in comments. Future work should include surveys or interviews to dive deeper into the user motivations and to critically assess our findings. This could include, for example, studies on the users who use scientific URLs in comments. Automatic approaches to the identification of motivations, e.g., via phrase detection, are needed to conduct similar studies at scale. However, as our exploratory study has shown, comments are less structured and less homogenous than citations and quotes in scholarly publications, often providing no context at all, which makes automatic processing of comments difficult.

**ACKNOWLEDGEMENTS**

The authors thank Altmetric.com for providing this study's data free of charge for research purposes. We would like to thank Arslan Zafar for dataset preparation and Lara Bieler who assisted in the third round of coding comments.